\documentclass[a4paper,fleqn]{cas-dc}
\usepackage[T1]{fontenc}

\usepackage{cite}
\usepackage{multirow}
\usepackage{amsmath,amssymb,amsfonts}
\usepackage{mathrsfs}
\usepackage{algorithmicx}
\usepackage{algpseudocode}
\usepackage[linesnumbered,ruled,vlined]{algorithm2e}
\usepackage{soul}
\usepackage{graphicx}
\usepackage{subfig}
\usepackage{textcomp}
\usepackage{xcolor}
\usepackage[numbers]{natbib}
\def\tsc#1{\csdef{#1}{\textsc{\lowercase{#1}}\xspace}}
\tsc{WGM}
\tsc{QE}
\tsc{EP}
\tsc{PMS}
\tsc{BEC}
\tsc{DE}

\begin{document}
\shorttitle{ENCRUST }
\shortauthors{G. Kuldeep \& Q. Zhang}

\title [mode = title]{Design and Prototype of a Unified Framework for Error-robust Compression and Encryption in IoT}                      

\tnotetext[1]{This work is supported by Innovation Fund Denmark (Grant no.8057-00059B)  and DIGIT center Aarhus University.}


\author{Gajraj Kuldeep}
\ead{gkuldeep@ece.au.dk}

\author{Qi Zhang}
\ead{qz@ece.au.dk}

\address{DIGIT, Department of Electrical and Computer Engineering \\ Aarhus University, Denmark
}



\begin{abstract}
	The Internet of Things (IoT) relies on resource-constrained devices for data acquisition, but the vast amount of data generated and security concerns present challenges for efficient data handling and confidentiality. Conventional techniques for data compression and secrecy often lack energy efficiency for these devices. Compressive sensing has the potential to compress data and maintain secrecy, but many solutions do not address the issue of packet loss or errors caused by unreliable wireless channels. To address these issues, we have developed the ENCRUST scheme, which combines compression, secrecy, and error recovery. In this paper, we present a prototype of ENCRUST that uses energy-efficient operations, as well as a lighter variant called L-ENCRUST. We also perform security analysis and compare the performance of ENCRUST and L-ENCRUST with a state-of-the-art solution in terms of memory, encryption time, and energy consumption on a resource-constrained TelosB mote. Our results show that both ENCRUST and L-ENCRUST outperform the state-of-the-art solution in these metrics.
\end{abstract}

%

\begin{keywords}
compressive sensing \sep sensing matrix generation \sep energy efficiency \sep joint compression and encryption \sep IEEE 802.15.4 \sep information security
\end{keywords}

\maketitle

\section{Introduction}
The internet of things (IoT) has enabled numerous applications and services in different verticals such as industrial control and automation, smart city, smart home, E-health, and many others.  The massive increase in IoT devices and continuous sensing has led to exponential growth in the data. According to Statista \cite{dataV}, there will be 79.4 Zetabytes of data in 2025 and  this will significantly increase network traffic. In most of the deployment scenarios, these IoT devices are resource-constrained \cite{intr1}.  Additionally, the confidentiality of data is paramount because data can be exploited to extract vital or private information. In summary, a sustainable IoT ecosystem needs novel solutions for data compression, error correction, and data confidentiality in resource-constrained devices.

\begin{figure}[htbp]
	\centering
	\includegraphics [width=1\linewidth]{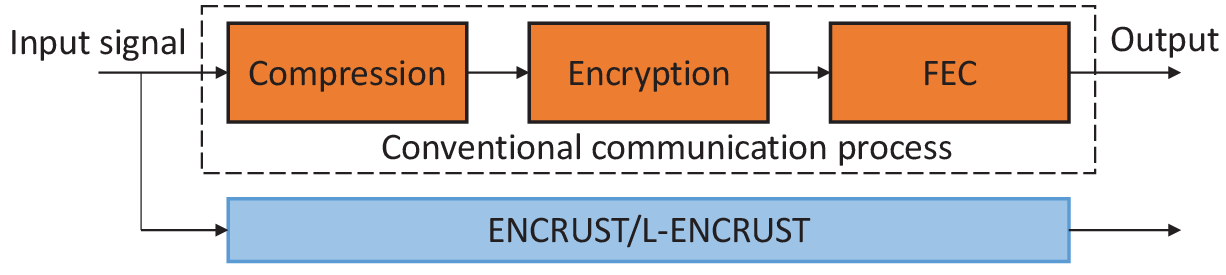}
	\caption{Conventional communication system vs unified framework of the ENCRUST and L-ENCRUST.}
	\label{blockDiag}
\end{figure}

In the conventional communication system, data compression, forward error correction, and information secrecy are achieved using three separate schemes, respectively as shown in Fig. \ref{blockDiag}. For example, compression can be achieved using discrete cosine transform (DCT) or wavelet transform followed by entropy coding  and  forward error-correcting codes (FEC) such as Reed-Solomon codes, low-density parity-check codes, etc, are used for forward error correction \cite{dataCom, dataError}. Information secrecy can be achieved by incorporating an encryption algorithm such as  advanced encryption standard (AES) \cite{ws}.   However, these conventional methods  are not energy-efficient when implemented in  resource-constrained devices.  It has been  empirically shown that usage of conventional data compression methods such as wavelet transform and DCT if implemented inappropriately in resource-constrained IoT devices could jeopardize energy efficiency as compared to the uncompressed data \cite{ECGWBSN, WSNC}.  Besides, the existing data encryption schemes are susceptible to channel errors, which is referred to as sensitivity encryption, as channel errors can result in decryption failure or low-quality decrypted data \cite{FEC1}.  One way to protect encrypted data from channel errors is to apply FEC with high error correction capability. The drawbacks of FECs in resource-constrained devices are  increased processing complexity and high energy consumption. Because higher error correction capability is often realized through longer codewords and lower coding rates (i.e., higher redundancy overhead).  Therefore, it is highly desirable to have an energy-efficient scheme that can achieve compression, information secrecy, and error recovery in one go.

In our recent work \cite{ENCRUST_IA}, the theoretical foundation of a novel scheme, efficient secure error-robust (ENCRUST), is proposed that can perform data compression, encryption, and error recovery within one single processing module at  IoT devices.  The encoding process of the ENCRUST is a simple matrix operation that is extremely beneficial for resource-constrained IoT devices. The decoding of the ENCRUST is composed of two $l_1$ minimization processes, the first one for error recovery and the second one for signal reconstruction. 

In this article, we implement a prototype of the ENCRUST scheme in  resource-constrained TelosB mote. Using linear feedback shift registers and lagged Fibonacci generators, energy-efficient construction methods for error recovery matrix and compression matrix are explored. We also design a new lightweight variant of ENCRUST, referred to as L-ENCRUST.   Furthermore, we compare the energy consumption of a state-of-the-art (SoA) conventional solution, ENCRUST, and L-ENCRUST using a real-life electrocardiogram (ECG) dataset.

\subsection{Main Contributions}
Our main contributions in this article can be summarized as follows.
\begin{itemize}
	\item We propose energy-efficient methods to construct error recovery matrix and compression matrix. A new energy-efficient lightweight scheme, L-ENCRUST, is realized by changing only one column of the error recovery matrix, instead of an entire matrix. In this way, only $L\times d$ operations are needed for constructing the error recovery matrix in L-ENCRUST instead of $L\times M$ operations in ENCRUST where $d<M$.
	
	\item Security analysis of the  ENCRUST and L-ENCRUST schemes is performed. This shows that both  ENCRUST and L-ENCRUST are resistant against ciphertext-only attack, known-plaintext attack, and chosen-plaintext attack.
	
	\item We study the error recovery performance of the ENCRUST and L-ENCRUST by designing simulation based on the physical layer of IEEE 802.15.4 standard with offset phase shift keying modulation and additive white Gaussian channel noise. Comparing with the SoA solution, both ENCRUST and L-ENCRUST achieve better error recovery performance.
	
	\item We design and implement prototypes for the ENCRUST and L-ENCRUST in  resource-constrained TelosB mote. We carry out a series of experiments using the prototypes to measure the memory footprint, encryption time and energy consumption of ENCRUST and L-ENCRUST at TelosB mote, and compare them with those of the existing solution. Clearly performance gains have been observed, for example, ENCRUST and L-ENCRUST can bring a reduction of $12\%$ and $26\%$ in the total energy consumption, respectively, compared with the SoA solution. 
	
\end{itemize}


\subsection{Related Work}
Compressive sensing (CS) allows sampling of a sparse or approximately sparse signal below the Nyquist sampling rate. It is a joint signal acquisition and compression method\cite{bib:EJT,bib:DLD}. CS has been widely applied in many  applications such as wireless communication, image processing, magnetic resonance
imaging, remote sensing imaging, information secrecy etc. \cite{IoTA,SWC}. Here we mainly present the related work that applying CS for joint compression and information secrecy, and CS for error recovery. 

Y. Rachlin et al. \cite{Rachlin}  propose that CS-based joint compression and information secrecy are achievable, if the sensing matrix is Gaussian distributed and changed for each sensing \cite{Rachlin}. In our recent work\cite{OurP}, it is shown that CS-based schemes are vulnerable to ciphertext attacks. However, such attacks are not possible for constant energy signals. CS-based joint compression and information secrecy scheme for constant energy signals is proposed in \cite{prefect,ourEnc}. CS has also shown potential for joint compression and  multi-class encryption \cite{mpcc}. Confidentiality-preserving compressed acquisition is proposed for multimedia using CS for resource-constrained devices\cite{CSN1}. Privacy-preserving CS scheme for image compression and encryption  is proposed in \cite{CSN2}, which incorporates a non-linear operation  in the CS encoding process to provide privacy. Joint encryption and compression scheme for audio signal using CS is proposed in \cite{CSN3}. Authors \cite{CSN3} first construct a sparse frame using discrete cosine transform then use one-time CS to provide information secrecy.

CS has also been applied to error correction. CS-based error correction on the Nyquist sampled data is introduced in \cite{ec1, ec2}. These error correction methods are applied  on $K$-sparse signals. Noise-resistance CS-based scheme is presented to remove the effect of measurement noise \cite{bib:LC1}. Error correction based on Fourier CS and projective geometry has been studied in \cite{bib:LC2}. Dense error correction for face images using $l_1$ minimization is explored in \cite{bib:LC3}.  Our work is different from the above mentioned schemes because the ENCRUST and L-ENCRUST provide compression, error recovery, and information secrecy.

Joint compression and error correction using arithmetic codes and turbo codes was attempted by E. Magli et al. \cite{bib:CEDMagli}.  They designed two algorithms for joint source, channel coding, and secrecy. These algorithms are designed using the arithmetic codes and turbo codes on the Nyquist sample data. Additionally,  these algorithms provide a weak sense of security because the turbo code-based algorithm provides secrecy by scrambling, and the arithmetic code-based algorithm uses randomized coders to provide secrecy \cite{ws}. The ENCRUST and L-ENCRUST provide computational secrecy, and  are both resistant against various cryptographic attacks such as  ciphertext-only attack, chosen-plaintext attack, and known-plaintext attack. Our  proposed schemes are dynamic, and one can change the desired compression ratio and error recovery capability depending on the channel conditions.

Prototype of CS-based compression is attempted in \cite{WBSN}. Authors \cite{WBSN} compare traditional compression methods using wavelet transform and Huffman coding with CS-based compression. Their findings are that the CS-based compression scheme performs better in energy efficiency than the conventional wavelet-based compression.  And it is very challenging to implement energy-efficient sensing matrix generation. Therefore, the authors use a fixed sensing matrix constructed using ones and zeros. Nevertheless, the aim of \cite{WBSN} is to design a CS-based compression scheme.  Therefore, error recovery and information secrecy are not considered in their work.

Prototype of the energy concealment scheme for joint compression and information secrecy is  presented in \cite{EC}. The proposed solution reduces the processing time to  the range of $80$ and $109$ milliseconds depending on the size of CS measurements for signal length 256. It is shown that the energy concealment scheme performs better in terms of energy efficiency as compared to AES. However, the energy concealment scheme does not incorporate  error recovery mechanism.

The first theoretical attempt to combine compression, information secrecy, and error recovery, referred to as ENCRUST, is presented in \cite{ENCRUST_IA}. In \cite{ENCRUST_IA}, it is shown that the compression and error recovery can be incorporated in the encoding processing using projection method, and information secrecy can be achieved by changing either the compression matrix or error recovery matrix for each sensing.

In this paper, we aim to develop a working prototype of  the ENCRUST scheme in TelosB mote and compare it with the state-of-the-art solution in terms of energy efficiency, memory storage, and processing time.  We design a lightweight version of the ENCRUST, L-ENCRUST, in which compression matrix and error recovery matrix are kept fixed and secrecy is provided by augmenting the column of the fixed error recovery matrix with a random vector.  

\begin{figure*}[htbp]
	\centering
	\includegraphics [width=.8\linewidth]{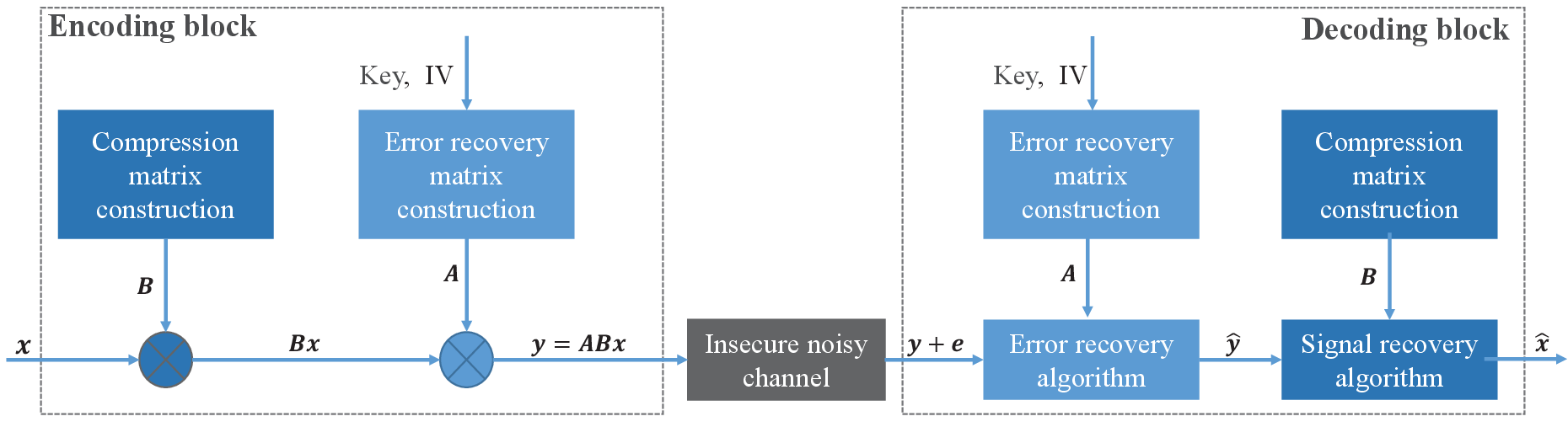}
	
	\caption{Framework of the ENCRUST scheme}
	\label{ECF}
\end{figure*}

\subsection{Organization and Notations}

The paper is organized as follows. In Section II the theoretical background  of  compressive sensing and the ENCRUST scheme  is presented. Section III presents the lightweight scheme, L-ENCRUST. In Section IV construction of the compression matrix and error recovery matrix is explained. Section V contains security analysis for the ENCRUST and L-ENCRUST. Section VI presents the performance evaluation based on the experimental results using the implemented prototype for the ENCRUST and L-ENCRUST, as well as the conventional solution. Finally, Section VII concludes the paper. 

\textit{Notations:} In this paper, all the boldface uppercase, e.g., $\mathbf{ X}$, and all the boldface lowercase, e.g., $\mathbf{ x}$, letters represent matrices and vectors, respectively. $\mathbf{x}^T$ is transpose of $\mathbf{x}$, similarly $\mathbf{ X}^T$ is transpose of $\mathbf{ X}$. The italic letters represent variables.  $l_p$ norm of a vector $\mathbf{x}$ is represented as $(\sum_{i=1}^{N}|x_i|^p)^{\frac{1}{p}}$. Symbol $\bigotimes$ represents a matrix and vector multiplication. Symbols $>>$ and $<<$ are left shift and right shift operations, respectively. Symbol $\oplus$ represents xor operation.

\section{Theoretical Background }
In this section, first compressive sensing basics for joint compression and information secrecy are  presented,  then the ENCRUST fundamentals are explained.
\subsection{Compressive Sensing}
Let $\mathbf{x}\in \mathbb{R}^N$ be a signal of length $N$. It can be either $K$-sparse signal in the canonical form, i.e., $||\mathbf{x}||_0=K$ or approximately sparse signal, also referred to as compressible signal. For an approximately sparse signal there exist a transform $\mathbf{\Psi}$ such that the most of signal information is contained in the $K$  coefficients of the signal transformation, $\mathbf{\theta=\Psi x}$.
Joint compression and information secrecy in CS can be achieved by taking random linear measurements using a Gaussian distributed sensing matrix, $\mathbf{\Phi}_i\in \mathbf{R}^{M\times N}$, where $M<N$.  Ciphertext, $\mathbf{y}_i$, for a plaintext $\mathbf{ x}_i$ is given as,
\begin{eqnarray}
\mathbf{y}_i=\mathbf{\Phi}_i \mathbf{x}_i. \label{csq1}
\end{eqnarray}
Rachlin et al. \cite{Rachlin} demonstrate that Eq. \ref{csq1} is  computationally secure if sensing matrix is used only once and its entries are Gaussian distributed.    
The plaintext, $\mathbf{x}_i$, can be recovered using convex optimization, if the signal satisfies the sparsity constraint and the sensing matrix satisfies the restricted isometric property (RIP)~\cite{bib:rip}. The optimization problem \cite{bib:EJT,bib:DLD} to recover plaintext is given as,
\begin{eqnarray}
\mathbf{\hat{x}}_i=\arg \min \limits_{\mathbf{x} \in \mathbb{C}^{N}} \mathbf{||x}||_1, \text{       s.t.   } \mathbf{y}_i=\mathbf{\Phi}_i \mathbf{x}.  
\end{eqnarray} 

If the signal is approximately sparse, then the optimization problem \cite{sparseD,sparseD1} becomes,
\begin{eqnarray}
\mathbf{\hat{\theta}}_i=\arg \min \limits_{\mathbf{\theta} \in \mathbb{C}^{N}} \mathbf{||\theta}||_1, \text{       s.t.   } \mathbf{y}_i=\mathbf{\Phi}_i\mathbf{\Psi}\mathbf{\theta}.  
\end{eqnarray}

\subsection{ENCRUST}
In this subsection, fundamentals of the ENCRUST \cite{ENCRUST_IA} scheme are explained. In ENCRUST, simultaneous compression and error recovery  are achieved by taking random linear measurements using a compression matrix, $\mathbf{B}\in \mathbf{R}^{M\times N}$ and an error recovery matrix, $\mathbf{A}\in \mathbf{R}^{L\times M}$, where $M<N$ and $M<L$.  In ENCRUST the measurement vector, $\mathbf{y}$, is given as,
\begin{equation}
\mathbf{y=AB x}. \label{csb}
\end{equation}
The signal, $\mathbf{x}$, can be recovered using convex optimization, if the signal satisfies the sparsity constraint and the compression matrix and error recovery matrix satisfy the RIP  \cite{ENCRUST_IA,bib:rip}. {In this paper, the error recovery matrix and the compression matrix are constructed using binary random number generator, and therefore these matrices satisfy the RIP.}

If the received measurement vector through a communication channel is corrupted with error vector $\mathbf{e}\in\mathbb{R}^L$ and $||\mathbf{e}||_0=\rho_0$ such that $L-M$ is in the order of $\rho_0log(L/\rho_0)$. The received measurement vector is given as,
\begin{equation}
\mathbf{y}_{rx}=\mathbf{y} + \mathbf{e}. \label{rre}
\end{equation} 

To estimate channel error,  a matrix $\mathbf{P}=\mathbf{ I}-(\mathbf{A}(\mathbf{A}^T \mathbf{A})^{-1}\mathbf{A}^T)$ can be constructed such that $\mathbf{P} \mathbf{A}=\mathbf{0}$. After multiplying $\mathbf{y}_{rx}$ in Eq. \ref{rre} with $\mathbf{P}$, the error projection $\mathbf{ep}$ is given as,
\begin{equation}
\label{ppr}
\begin{aligned}
\mathbf{ep} &= \mathbf{P} \mathbf{y}_{rx},\\
&= \mathbf{P} \mathbf{e}. 
\end{aligned} 
\end{equation} 
The error vector, $\mathbf{e}$, in Eq. \ref{rre} can be estimated  using $l_1$ minimization and the optimization problem  can be formulated as,
\begin{equation}
\mathbf{\hat{e}}=\arg \min \limits_{\mathbf{e} \in \mathbb{R}^{N}} \mathbf{||e}||_1, \textit{         s.t.     } \mathbf{ep}=\mathbf{P} \mathbf{e}. \label{l1error} 
\end{equation} 

{The detailed proof for the error recovery using orthogonal projection matrix, $\mathbf{ P}$, is given in Section III.A of \cite{ENCRUST_IA}.} The estimated error vector, $\mathbf{\hat{e}}$, is subtracted from $\mathbf{y}_{rx}$ and the estimated measurement vector is given as,
\begin{equation}
\mathbf{\hat{y}}=\mathbf{y}_{rx}-\mathbf{\hat{e}}.
\end{equation}

The signal $\mathbf{x}$ can be reconstructed by performing $l_1$ minimization and the optimization problem  can be formulated as, 
\begin{equation}\label{l1signal}
\mathbf{\hat{x}}=\arg \min \limits_{\mathbf{x} \in \mathbb{R}^{N}} \mathbf{||x}||_1, \textit{         s.t.     } \mathbf{A}^T\mathbf{\hat{y}}=\mathbf{A}^T\mathbf{AB} \mathbf{x}.  
\end{equation} 
It can be observed that to reconstruct the signal, $\mathbf{x}$,  $l_1$ minimization is performed twice. First using the orthogonal projection matrix $\mathbf{P}$ to recover the error vector from the received measurement vector, $\mathbf{y}_{rx}$, using Eq. \ref{l1error}. After the estimation of the error vector, $\mathbf{e}$, $l_1$ minimization is performed to reconstruct the encoded signal using Eq. \ref{l1signal}. The design parameters of the ENCRUST are the number of significant coefficients for an approximately sparse signal, $K$ and the error correction capability $\rho_0$. Hence, in ENCRUST the dimension of the measurement vector can be expressed as,
\begin{equation} \label{Ldim}
\begin{aligned}
L &=\alpha_1 K+\alpha_2 \rho_0, \\
& =M+\alpha_2 \rho_0,
\end{aligned}
\end{equation}
where $K$ is the sparsity of the signal, $\rho_0$ is the error correction capability, and  $\alpha_1\ge2$ and $\alpha_2 \ge2$ are constants. The parameter $L$ can be tuned according to the channel conditions, because error correction capability $\rho_0$ is given as $\frac{L-M}{\alpha_2}$ for a particular $L$.

The framework of the ENCRUST scheme is shown in Fig. \ref{ECF}. This framework is equivalent to a symmetric key encryption algorithm, because the secret key is shared between the receiver and transmitter. The encryption of a plaintext block takes place as follows: First, compression matrix is applied on the signal for compression. After that error recovery matrix is applied on the compressed signal for error recovery and information secrecy. Finally, the ciphertext is transmitted through insecure noisy channel. The compression matrix is fixed and the error recovery  matrix is constructed from the pseudo-random sequence generated using the secret key and initialization vector (IV). In the decryption part, the same error recovery matrix is generated using the knowledge of the secret key and IV at the receiver, based on which the channel error can be estimated using optimization algorithms, such as Basis Pursuit. After subtracting the estimated channel error from received ciphertext, signal can be reconstructed again using optimization algorihtm as in Eq. \ref{l1signal}.

In \cite{ENCRUST_IA}, it is shown that from Eq. \ref{csb} information secrecy can be achieved either by changing the matrix $\mathbf{A}$  or by changing matrix $\mathbf{B}$  for each sensing. If one of the matrices is changed for each sensing then the achieved information secrecy is equivalent to one-time CS-based security.  In the next section, a variant of ENCRUST is proposed to provide information secrecy  without changing matrix $\mathbf{A}$ and $\mathbf{B}$.  

\section{Lightweight ENCRUST} \label{llss}
The ENCRUST is designed to provide compression and error recovery using  compression matrix and error recovery matrix, respectively. The information secrecy is achieved through changing one of the matrices for each sensing, which can be an energy consuming operation in resource-constrained IoT devices.  The notion of secrecy here is that an adversary will not be able to reconstruct the signal intended
for a legitimate receiver. In other words, an adversary’s probability of reconstruction of the original signal is low. 

This new variant of ENCRUST scheme achieves information secrecy by augmenting the error recovery matrix. Let  $\mathbf{A}$ and $\mathbf{B}$  be fixed and random matrices. The augmented error recovery matrix is given as,
\begin{equation} \label{n1}
\mathbf{A}_u=[\mathbf{r} \hspace{2mm} \mathbf{A}]_{L\times (M+1)},
\end{equation}
where $\mathbf{r}\in \mathbb{R}^L$ is a random vector and its entries are drawn from  uniform distribution. The augmented compression matrix is given as,
\begin{equation}\label{n2}
\mathbf{ B}_u=\begin{bmatrix}
1& \mathbf{ 0}_{1\times N} \\ 
\mathbf{ 0}_{M\times 1}& \mathbf{ B} 
\end{bmatrix}_{(M+1)\times (N+1)}.
\end{equation} 
The transmitted signal vector using augmented error recovery matrix and augmented compression matrix for a signal $\mathbf{ x}$ is given as,
\begin{equation} \label{ENCV}
\mathbf{ y}=\mathbf{A}_u\mathbf{ B}_u\begin{bmatrix}
c\\\mathbf{ x}
\end{bmatrix},
\end{equation}  
where $c$ is an arbitrary constant. In this paper, we take $c=1$. 

The received signal vector is corrupted with error and given as,
\begin{equation}
\mathbf{y}_{rx}=\mathbf{y} + \mathbf{e}. \label{arv}
\end{equation} 
Orthogonal projection matrix, $\mathbf{P}$, is constructed using $\mathbf{A}_u$ and error vector is estimated using Eq. \ref{l1error}. Augmented signal $\begin{bmatrix}
c\\\mathbf{ x}
\end{bmatrix}$ is recovered using $\mathbf{A}_u$, $\mathbf{B}_u$, $\mathbf{\hat{e}}$, and Eq. \ref{l1signal} and represented as $\mathbf{\hat{x}}_u$. This scheme is referred as L-ENCRUST in the paper.
In the next section, we present the methods to construct the compression matrix, error recovery matrix, and random vector.
\section{Construction of Compression Matrix and Error Recovery Matrix} \label{matDes}

Constructions of the compression matrix and error recovery matrix depend on the functionalities achieved using Eq. \ref{csb}. Suppose only compression and error recovery are required, then the product of the fixed matrices $\mathbf{A}$ and $\mathbf{ B}$ can be precomputed and stored. However, to achieve information secrecy, one of the matrices should be changed for each sensing. Additionally, to store a big matrix in resource-constrained IoT devices may be either infeasible or inefficient.  Therefore, it is necessary to explore  on the fly generation of the compression matrix and error recovery matrix. {In this section, we propose two algorithms to construct the error recovery matrix and the compression matrix using random numbers. The $\mathbf{ A}$ and $\mathbf{ B}$ matrix in L-ENCRUST are the error recovery matrix and compression matrix, respectively, and information secrecy in L-ENCRUST is provided by random vector, $\mathbf{r}$. In the ENCRUST,  matrix $\mathbf{ A}$  provides both error-recovery function and information secrecy, therefore it is required to be changed for every data block. Therefore, matrix $\mathbf{ A}$ in ENCRUST is constructed using binary matrix construction (Algorithm \ref{algoFmC}). On the other hand the matrix $\mathbf{ B}$ in ENCRUST is used only for compression and is constructed by sparse matrix construction (Algorithm \ref{algoSm}).}

\subsection{Sparse Matrix Construction} \label{smc}
In this subsection, we propose sparse matrix construction method using LFSR, which is more efficient compared to the binary matrix construction method. The pseudo-code to construct a run-time sparse matrix, $\mathbf{ \Phi}$, is given in Algorihtm \ref{algoSm}. In this method, each row can contain $d$ number of non-zero entries in the constructed matrix. Parameters $\text{shiftBits}$ and  $\text{lsbMask}$ are decided using the LFSR length and $N$. For the experiments in the paper,  $\text{shiftBits}$ and  $\text{lsbMask}$ are used as $8$ and $0$x$00FF$, respectively. The binary representation of the feedback polynomial and  initialization vector are  $\text{fp}$ and  $\text{iv}$, respectively.

\begin{figure}[htbp]
	\centering
	\includegraphics [width=.8\linewidth]{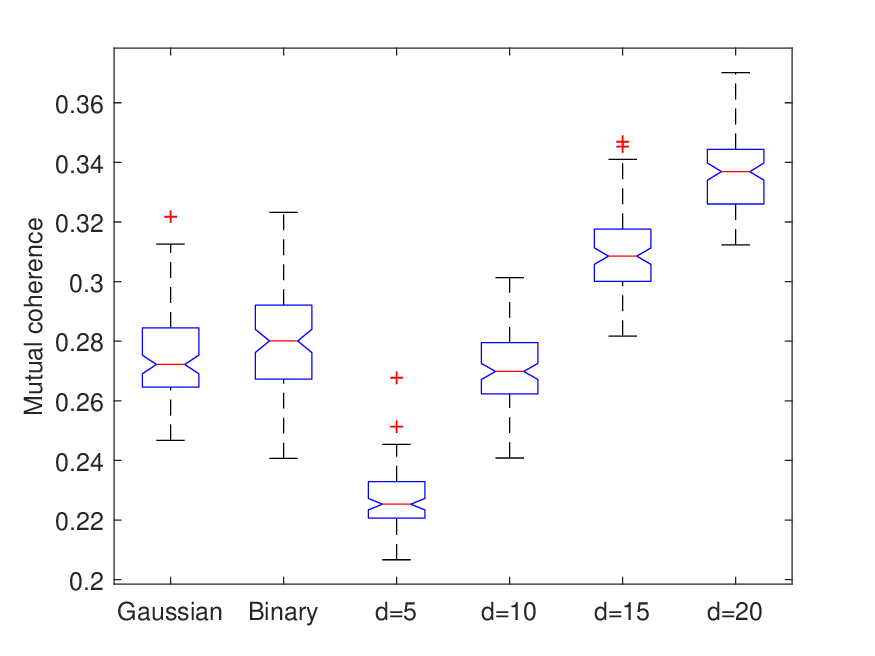}
	\caption{Illustration of mutual coherence values for a Gaussian distributed matrix, a binary matrix, and sparse matrices for various values of $d$ and $N=256$. }
	\label{mutCoh}
\end{figure}

\begin{algorithm}[h]
	\SetKwInput{KwInput}{Input}                
	\SetKwInput{KwOutput}{Output}              
	
	\DontPrintSemicolon
	\vspace{.3cm}
	\KwInput{$\text{iv}$, $\text{fp}$, $\text{shiftBits}$, $\text{lsbMask}$, $d$, $M$, $N$ }
	
	
	$\mathbf{ \Phi}=0$\;
	$\text{sr}=\text{iv}$\;
	\For{$i=1$ to $M$}{
		
		\For{$j=1$ to $d$}{
			\If{(MSB$(\text{sr})$)}
			{	$\text{sr}=\text{sr} \oplus \text{fp}$\;
				$\text{jIndex}=(\text{sr}>>\text{shiftBits})\oplus (\text{sr}$ \& $\text{lsbMask})$\;
				$\phi(i,\text{jIndex})=\phi(i,\text{jIndex})-1$\;	
			}
			\Else{
				$\text{jIndex}=(\text{sr}>>\text{shiftBits})\oplus (\text{sr}$ \& $\text{lsbMask})$\;
				$\phi(i,\text{jIndex})=\phi(i,\text{jIndex})+1$\;	
			}
			
			$\text{sr}=\text{sr}<<1 $\;	
			
		}
	}
	
	\KwOutput{$\mathbf{ \Phi}$}
	\caption{Sparse matrix construction }\label{algoSm}
\end{algorithm}
We study the mutual coherence property for the proposed matrix constructions.  In compressive sensing, mutual coherence is studied to guarantee reconstruction of the sensed signal  from the measurement vector \cite{MCD}.
Mutual coherence between a matrix $\mathbf{ \Phi}$ and sparsifying basis $\mathbf{\Psi}$ is given as,
\begin{equation}
\mu(\mathbf{ \Phi},\mathbf{\Psi})=\max \limits_{ 1 \le i,j \le N}\frac{| \mathbf{\phi}^T_i\mathbf{\psi}_j|}{||\mathbf{\phi}_i||_2 ||\mathbf{\psi}_j||_2}
\end{equation}
where $\mathbf{\phi}_i$ is the $i^\text{th}$ column vector of matrix, $\mathbf{\Phi}$, and $\mathbf{\psi}_j$ is the $j^\text{th}$ column vector of matrix, $\mathbf{\Psi}$. Values of the mutual coherence for matrices described in Subsection  \ref{bmc} and \ref{smc} are compared with Gaussian matrix in Fig. \ref{mutCoh}. The range of $\mu$ for Gaussian/binary matrix and DCT sparsifying matrix is from $0.24$ to $0.32$.   From Fig. \ref{mutCoh}, it can be observed that the mutual coherence values are equivalent to the Gaussian matrix for matrices constructed using Algorihtm \ref{algoSm} for $d=10$ to $d=15$. 
\subsection{Binary  Matrix Construction}  \label{bmc}
Real-time generation of a binary matrix can be realized using linear feedback shift register (LFSR).  The pseudo-code to construct a run-time binary matrix, $\mathbf{\Phi}$, is given in Algorihtm \ref{algoFmC}. A matrix of size $M\times N$ is constructed depending on the feedback polynomial and initialization vector. The binary representation of the feedback polynomial and $i^\text{th}$ initialization vector are  $\text{fp}$ and  $\text{iv}_i$, respectively. In the prototype, we use 16-bit LFSR with the primitive polynomial, $x^{16}+x^{14}+x^{13}+x^{11}+1$, to construct a binary matrix.

\begin{algorithm}[h]
	\SetKwInput{KwInput}{Input}                
	\SetKwInput{KwOutput}{Output}              
	
	\DontPrintSemicolon
	\vspace{.3cm}
	\KwInput{$\text{iv}_i$, $\text{fp}$, $M$, $N$ }
	

	\For{$i=1$ to $M$}{
		$\text{sr}=\text{iv}_i$\;
		\For{$j=1$ to $N$}{
			\If{(MSB$(\text{sr})$)}
			{	$\text{sr}=(\text{sr}<<1) \oplus \text{fp}$\;
				$\phi(i,j)=1$\;

			}
			\Else{
				$\text{sr}=\text{sr}<<1 $\;
				$\phi(i,j)=-1$\;	
				
			}
		}
	}
	
	\KwOutput{$\mathbf{ \Phi}$}
	\caption{Binary matrix construction }\label{algoFmC}
\end{algorithm}

\subsection{Random Number Construction } \label{rnc}
\begin{figure}[htbp]
	\centering
	\includegraphics [width=.6\linewidth]{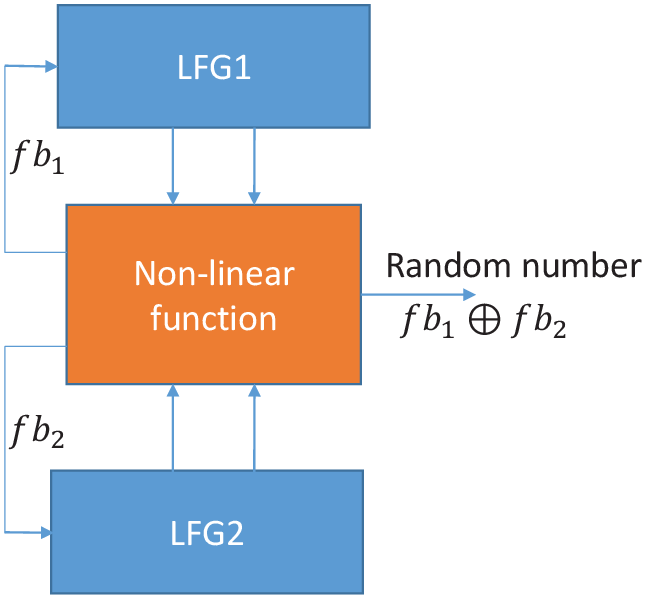}
	\caption{ Random number construction using LFG trinomials and Non-linear function. }
	\label{rng}
\end{figure}

In this subsection, construction of random numbers is proposed using lagged Fibonacci generators (LFG) \cite{lfg}. 
For efficient construction of random numbers, we use LFG trinomials. LFG trinomials $x^7+x^3+1$ and $x^5+x^2+1$ are used as LFG1 and LFG2, respectively, for random number construction as shown in Fig. \ref{rng}.  
Non-linear function is used to calculate feedback values to LFG1 and LFG2. Feedback values to LFG1 and LFG2 are given as,
\begin{equation}
\begin{aligned}
fb_1=(\text{LFG1 }[7]\oplus(\text{LFG2 }[2]<<3))\\ \oplus(\text{LFG1 }[3]\oplus(\text{LFG2 }[5]<<5))
\end{aligned}
\end{equation}
and
\begin{equation}
\begin{aligned}
fb_2=(\text{LFG2 }[5]\oplus(\text{LFG1 }[3]<<2))\\ \oplus(\text{LFG2 }[2]\oplus(\text{LFG1 }[7]<<7)),
\end{aligned}
\end{equation}
respectively. Registers in LFG1 and LFG2 are represented using 16-bits. 

\section {Security Analysis}
In this section, we study different attack models applied on the  ENCRUST and L-ENCRUST schemes.  The considered attacks are ciphertext-only attack, known-plaintext attack, and chosen-plaintext attack. Kerckhoffs' principles \cite{ws} are followed to apply these attacks, which state that {a} cryptographic system should be secure even if an adversary knows everything about the cryptosystem except {the} key. For ENCRUST, the master key consists of the matrix $\mathbf{ B}$ and secret used to construct matrix $\mathbf{ A}$. Similarly for L-ENCRUST, the master key consists of the matrices $\mathbf{ A}$ and $\mathbf{ B}$, and secret used to construct the random vector, $\mathbf{r}$.  
\subsection{Security Analysis of ENCRUST} For a plaintext $\mathbf{ x}_i$ the ciphertext  $\mathbf{ y}_i$ using ENCRUST is given as,
\begin{equation}
\mathbf{y}_i=\mathbf{A}_i\mathbf{B x}_i, \label{encEn}
\end{equation}
where $\mathbf{B}$ is a fixed compression matrix constructed using Algorihtm \ref{algoSm} and $\mathbf{A}_i$ is an error recovery matrix for $i^{th}$ sensing and  is constructed using Algorihtm \ref{algoFmC}. Since $\mathbf{ B}$ is a fixed matrix, Eq. \ref{encEn} can be written as,
\begin{equation}
\mathbf{y}_i=\mathbf{A}_i\mathbf{s}_i, \label{encS}
\end{equation}
where $\mathbf{s}_i=\mathbf{ Bx}_i$. 
The received ciphertext is given as,
\begin{equation}
\mathbf{y}_{rx_i}=\mathbf{y}_i+\mathbf{e}_i, \label{RxencS}
\end{equation}
where $\mathbf{e}_i$ is channel error vector.
Eq. \ref{encS} can be considered as one-time CS-based  encryption scheme.
\subsubsection{Ciphertext-only Attack} To retrieve the plaintext from the received ciphertext, $\mathbf{y}_{rx_i}$, the adversary should first estimate the  error vector, $\mathbf{e}_i$, which requires knowledge of the error recovery matrix $\mathbf{A}_i$. Since the size of $\mathbf{ A}_i$ is $L\times M$, there are $2^{LM}$ possible error recovery matrices. The most efficient recovery algorithm's computational complexity is $O(N^{1.2})$ \cite{l1Min}. Therefore, the overall computational cost to apply brute force attack on ENCRUST is $2^{LM}O(N^{1.2})$. The ciphertext-only attack on the conventional CS-based encryption is presented in \cite{OurP} which exploits the inter-correlation among the encrypted blocks. However, this attack is not applicable to ENCRUST, because there are two matrix multiplications which remove the inter-correlation among the encrypted blocks and the decoding requires two $l_1$ minimization for plaintext information retrieval.

\subsubsection{Known-plaintext Attack and Chosen-plaintext Attack} In  known-plaintext attack, the adversary posses  pairs of plaintext and ciphertext. Then according to this information the adversary will try to find the error recovery matrix and ultimately get the key. In our recent study, we have shown in Section V.B of \cite{ourEnc} that the possible number of candidates is exponentially with plaintext length, which makes this attack infeasible.  Chosen-plaintext attack is possible on binary matrix, if binary matrix is used more than once, using superimposing sequences. However, in our case the error recovery matrix is changed for each sensing. Therefore, this attack is not possible.  Detailed security analysis of binary matrix based encryption using compressive sensing can be found in  our recent work \cite{ourEnc,EC}. 
\subsection{Security Analysis of L-ENCRUST} 
For encrypting a plaintext $\mathbf{ x}_i$, the ciphertext  $\mathbf{ y}_i$ using L-ENCRUST is given as,

\begin{equation} \label{LE}
\mathbf{ y}_i=\mathbf{A}_u\mathbf{ B}_u\begin{bmatrix}
c\\\mathbf{ x}_i
\end{bmatrix}.
\end{equation}  
Based on Eq. \ref{n1} and \ref{n2}, Eq. \ref{LE} can be written as,
\begin{equation} \label{n3}
\mathbf{ y}_i= \mathbf{A} \mathbf{B} \mathbf{x}_i+c\mathbf{ r}_i.
\end{equation}
To provide information secrecy in L-ENCRUST, random vector, $\mathbf{ r}_i$ is changed for each sensing. The entries of $\mathbf{ r}_i$ are taken from the method described in subsection \ref{rnc} and represented in 16 bits. We consider the following attack scenarios for L-ENCRUST.
\subsubsection{Ciphertext-only Attack} In this attack, it is assumed that the attacker has access of ciphertext.  To retrieve plaintext from ciphertext, the attacker requires to have the knowledge of $\mathbf{ A}$, $\mathbf{ B}$, and $\mathbf{ r}_i$.  The  entries of the random vector follow a uniform distribution,  therefore each entry of the random vector is equiprobable. 
To retrieve plaintext from the received ciphertext, $\mathbf{y}_{rx_i}$, the adversary should first estimate the  error vector, $\mathbf{e}_i$, which requires the knowledge of the random vector $\mathbf{r}_i$. Since the size of $\mathbf{ r}_i$ is $L\times 1$ and each entry of $r_i$ is represented in 16 bits, there are $2^{16L}$ possible random vectors. Therefore, it is computationally infeasible to remove the random vector from the ciphertext, even if the adversary has the knowledge of $\mathbf{ A}$ and $\mathbf{ B}$. 
\subsubsection{Known-plaintext and Chosen-plaintext Attack} In known-plaintext attack, the attacker posses  pairs of plaintext and its corresponding ciphertext with an aim is to retrieve key of random number generator. For this the attacker first tries to find matrices $\mathbf{ A}$ and $\mathbf{ B}$, then to retrieve the random vector and key. However, we show that it is impossible to retrieve the matrices  $\mathbf{ A}$ and $\mathbf{ B}$ from the plaintext and ciphertext pairs. The received ciphertext is denoted as $\mathbf{ z}_{i}$ for better readability and it is given as, 
\begin{equation} \label{ME}
\mathbf{ z}_{i}= \mathbf{A} \mathbf{B} \mathbf{x}_i+c\mathbf{ r}_i+\mathbf{ e}_i,
\end{equation}
where $\mathbf{ e}_i$ is channel error vector. Assuming that the adversary knows the $N$ plaintext ciphertext pairs of Eq. \ref{ME}. For $\mathbf{ e}_i=\mathbf{ 0}$, $\mathbf{ AB=H}$ and $c=1$, Eq. \ref{ME} can be given as,
\begin{equation} \label{ME1}
\mathbf{ z}_{i}= \mathbf{H}\mathbf{x}_i+\mathbf{ r}_i.
\end{equation}

\begin{figure}[htbp]
	\centering
	\subfloat[]{\includegraphics [width=.5\linewidth]{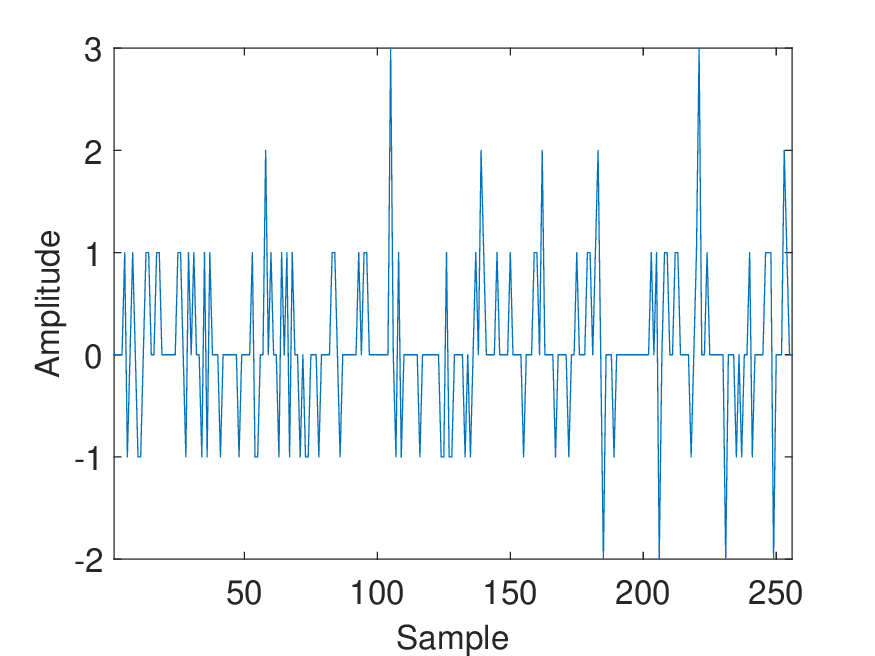}}
	\subfloat[]{\includegraphics [width=.5\linewidth]{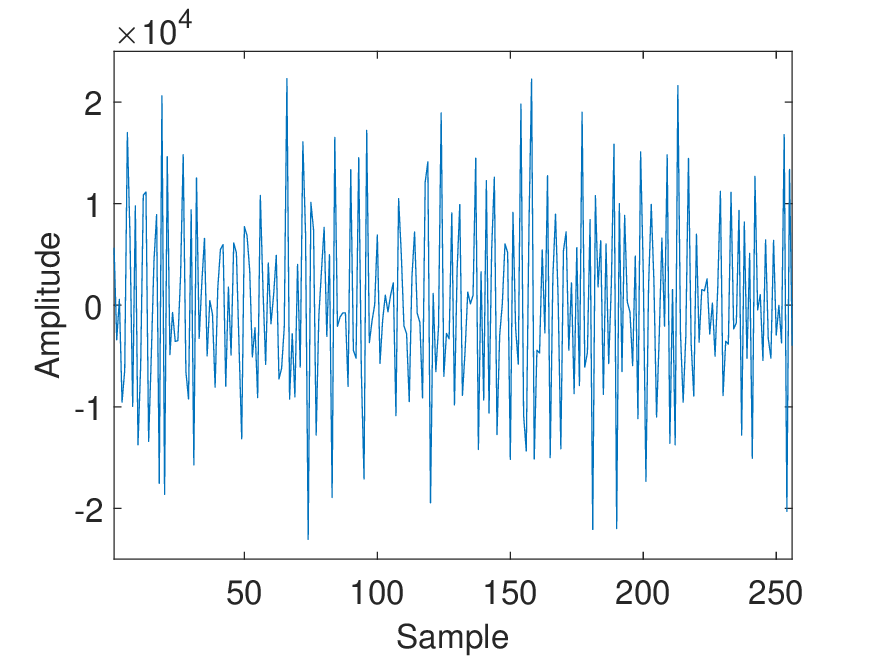}}
	\caption{(a) Original first row of  $\mathbf{ H}$ at the L-ENCRUST encoding. (b) Reconstructed first row of $\mathbf{ H}$ at the adversary using the known-plaintext attack.}
	\label{attackH}
\end{figure}

The adversary can construct a signal matrix $\mathbf{ X}$ as,
\begin{equation}
\mathbf{ X}=\begin{bmatrix}
\mathbf{ x}^T_1\\
\vdots \\
\mathbf{ x}^T_N\\
\end{bmatrix},
\end{equation}
and a vector $\mathbf{ d}$ as 
\begin{equation}
\mathbf{ d}=\begin{bmatrix}
{ z}^1_{1}\\
\vdots \\
{ z}^1_{N}\\
\end{bmatrix},
\end{equation}
where $z^1_i$ is a first value of the $i^{th}$ vector $\mathbf{ z}_{i}$. If inverse of the $\mathbf{ X}$ exists, then the adversary can try to estimate the first row of matrix $\mathbf{ H}$ by evaluating $\mathbf{ X}^{-1}  \mathbf{ d}$, which is given as,
\begin{equation} \label{ME2}
\mathbf{ X}^{-1}  \mathbf{ d}=\begin{bmatrix}
h_{11}\\
\vdots \\
h_{1N}
\end{bmatrix}+ \mathbf{ X}^{-1} \begin{bmatrix}
{ r}^1_{1}\\
\vdots \\
{ r}^1_{N}\\
\end{bmatrix},
\end{equation}
where $r^1_i$ is a first value of the $i^{th}$ vector $\mathbf{ r}_{i}$. From Eq. \ref{ME2}, it can be observed that the matrix $\mathbf{ H}$ cannot be recovered even when channel error vector is zero. We can also present the empirical evidence using the reconstruction error in the construction of matrix $\mathbf{ H}$. One example of original first row of the matrix $\mathbf{ H}$ and reconstructed using Eq. \ref{ME2} is shown in Fig. \ref{attackH}.

\begin{figure}[htbp]
	\centering
	\subfloat[\label{attackABa}]{\includegraphics [width=.5\linewidth]{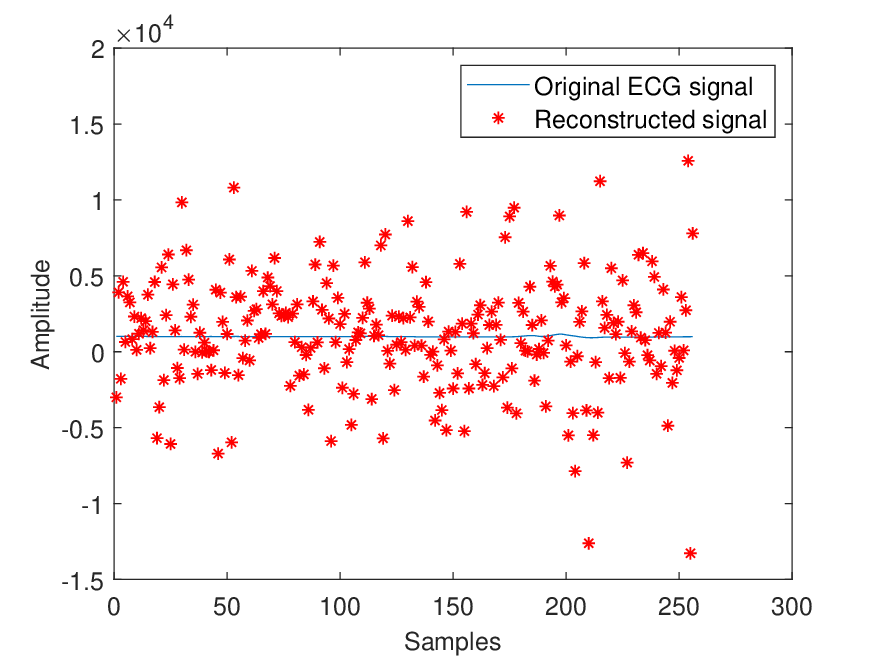}}
	\subfloat[\label{attackABb}]{\includegraphics [width=.5\linewidth]{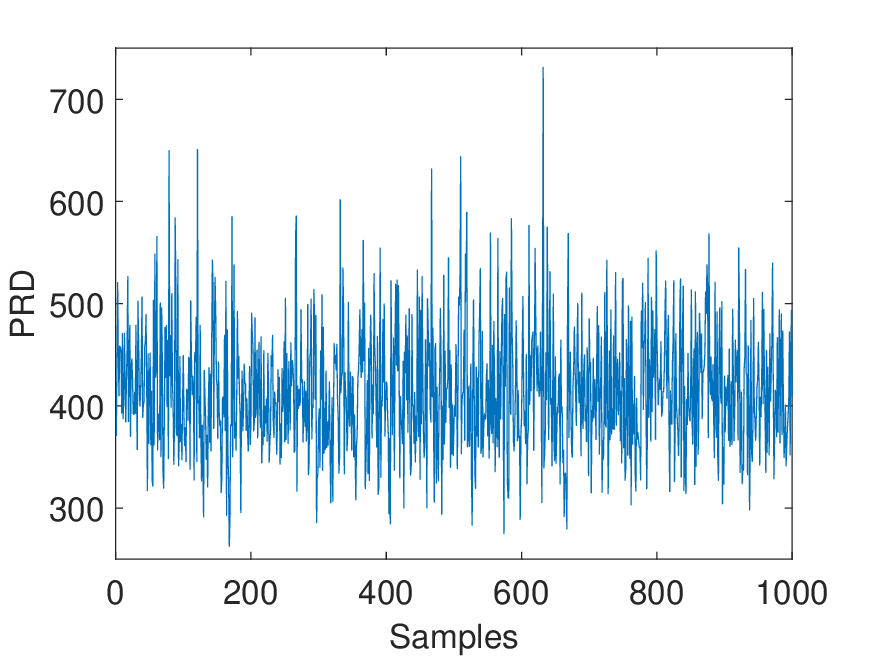}}
	\caption{Attack scenario on the L-ENCRUST when the adversary has the knowledge of $\mathbf{ A}$ and $\mathbf{ B}$ matrices   (a) Original and reconstructed signal at the adversary. (b) Reconstructed performance using PRD.}
	\label{attackAB}
\end{figure}

The chosen-plaintext attack is a more sophisticated attack, because in this scenario the adversary can choose plaintext. From Eq. \ref{ME2}, we know that to find matrix $\mathbf{ H}$, the knowledge of random vector is required. As random vector is changed for each sensing, the adversary can try to retrieve the key of the random number generator. However, the used random number generator is resistant to the cryptanalysis \cite{lfg}. For any plaintext, selected by the adversary, estimation of the entries of the matrix $\mathbf{ H}$ can be performed by Eq. \ref{ME2}. Suppose, the adversary learns the matrix $\mathbf{ H}$ for some combinations of plaintext. To remove the effect of random vector, the knowledge of error recovery matrix is required. But  finding of the $\mathbf{ A}$ and $\mathbf{ B}$ from $\mathbf{ H}$ is a matrix separation problem and there are infinitely many possibilities of  $\mathbf{ A}$ and $\mathbf{ B}$ for a given $\mathbf{ H}$. Therefore, it can be concluded that the knowledge of plaintext and ciphertext pairs does not help adversary in finding the matrices $\mathbf{ A}$ and $\mathbf{ B}$. We also demonstrate that an adversary is not capable to reconstruct signal even for the case that not only the adversary knows both matrices $\mathbf{ A}$ and $\mathbf{ B}$, but also no channel error vector is added. The adversary in this case can construct the augmented error recovery matrix by choosing an arbitrary random vector $\mathbf{ r}$. For this case, the reconstructed ECG signal by the adversary and original ECG signal are shown in Fig. \ref{attackABa}. The quality of the reconstructed signal by the adversary is measured using PRD as shown in Fig. \ref{attackABb}. The observed PRD values for one thousand samples vary from 262 to 731, which shows that reconstructed signal is by no means able to reflect the original signal. From the analysis presented in this subsection, it can be concluded that  ciphertext-only,  known-plaintext,  and chosen-plaintext attacks are not possible for the L-ENCRUST scheme.

\section{Performance Measurement } 
\subsection{Comparison Methods} \label{ComMe}
In this paper, we study the system-level performance in terms of energy consumption, processing time, storage, and error recovery   by comparing  the ENCRUST and L-ENCRUST with a state-of-the-art solution. The  implementations for different solutions studied  in this paper are described as follow.   

\subsubsection{Conventional solution} \label{ConHAH}
The data compression,  error recovery, and information secrecy are realized by three independent process, namely, Haar transform, Hamming codes, and AES, respectively. 
Data compression using Haar transform is considered to be energy-efficient because it requires only additions and subtractions. Data compression using Haar transform is performed simply by taking only average part of the signal and thresholding the difference part. Error recovery using Hamming based forward error correction with code rate $4/7$. Information secrecy is achieved using AES algorihtm, as it is the most widely used encryption algorihtm in resource-constrained devices. This combination for compression, error correction, and information secrecy is referred to as the state-of-the-art (SoA) solution in the paper. 

\subsubsection{ENCRUST} \label{SENCRUST}In this implementation, we consider the scheme described using Eq. \ref{csb} and only error recovery matrix is changed for each sensing to provide information secrecy. Error recovery matrix, $\mathbf{A}$, is constructed using Algorihtm \ref{algoFmC} as described in Subsection \ref{bmc} and $\text{iv}_i$ is taken from the random number generator as described in Subsection \ref{rnc}. Compression matrix, $\mathbf{ B}$, is constructed using Algorithm \ref{algoSm} with fixed $\text{iv}$ as described in Subsection \ref{smc}.

\subsubsection{L-ENCRUST } \label{SLENCRUST}In this implementation, we consider the scheme described using Eq. \ref{ENCV} and only error recovery matrix's first column is changed for each sensing to provide information secrecy. Both error recovery matrix, $\mathbf{A}$, and  compression matrix, $\mathbf{ B}$, are constructed using Algorithm \ref{algoSm} with fixed $\text{iv}$ as described in Subsection \ref{smc}. Care should be taken while constructing the error recovery matrix because it should be full rank. Vector $\mathbf{ r}_i$ is constructed using the method described in Subsection \ref{rnc}.

\subsection{Error Recovery Performance }
In this subsection, we compare the error recovery performance of the SoA, ENCRUST and L-ENCRUST.  To validate the performance of the ENCRUST and L-ENCRUST, we use MIT-BIH ECG database \cite{dbECG1,dbECG2}. This database contains 48 ECG records, each record is sampled using $360$ Hz and represented in 11-bit resolution. Reconstruction performance of ECG signals is measured using percentage root mean square difference (PRD)\cite{ARPD} between the original signal and the reconstructed signal, which is given as, 
\begin{equation}\label{prd}
\text{PRD}=(\sqrt{\frac{\sum_{i=1}^{N}|x_i-\hat{x}_i|^2}{\sum_{i=1}^{N}|x_i|^2}})100,
\end{equation}
where $x_i$ and $\hat{x}_i$ are $i^\text{th}$ samples of the original  and  reconstructed ECG signals, respectively. It is clear from Eq. \ref{prd} that lower PRD value means better signal reconstruction quality.

As ENCRUST and L-ENCRUST schemes use sparse sensing matrix, the design parameter $d$ is fixed by studying the mutual coherence and reconstruction performance.{There is a trade-off between the lower mutual coherence and reconstruction \cite{ECGWBSN}.  In \cite{ECGWBSN}, it is explained that signal to noise ratio vs number of non-zero elements in sensing matrix have trade-off shown in figure 4 of \cite{ECGWBSN}. For smaller values of $d$  we get lower value of mutual coherence but at same time it is not enough to capture the signal dynamics because there are many zeros in the compression matrix. This is the reason we take higher value of $d$. } From Fig. \ref{mutCoh}, we know that the parameter should be between $10$ and $15$ to have mutual coherence value similar to Gaussian and binary matrices.   We use Analog-to-Digital Converter (ADC) values of 100, 104, 111, 210, and 230 ECG records and perform simulations for $M=96$, $L=150$, and $N=256$. We study the reconstruction performance for ECG signals by considering various values of $d$. The PRD performance does not vary significantly in the range $10$ to $15$. In this paper, all simulations use the compression matrix $\mathbf{ B}$ constructed with $d=15$.

\begin{figure}[htbp]
	\centering
	\includegraphics [width=.8\linewidth]{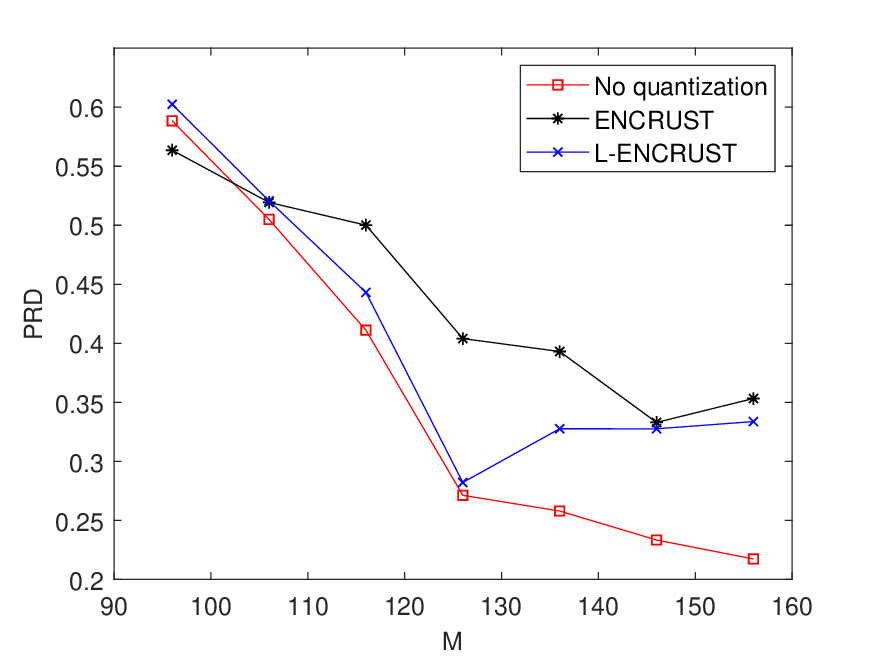}
	\caption{Effect of quantization on the ENCRUST and L-ENCRUST scheme for $N=256$, $L=168$, and various values of $M$.  }
	\label{quant}
\end{figure}

Note that compression and error recovery capabilities of the ENCRUST and L-ENCRUST are the same. In the prototype,  the ENCRUST and L-ENCRUST measurements are quantized to 16-bits. The received signal  with quantization error, $\mathbf{ e}_{q}$, and channel error vector, $\mathbf{ e}$, is given as,
\begin{equation}
\mathbf{y}_{rx}=\mathbf{y} + \mathbf{e}+ \mathbf{e}_q. 
\end{equation} 
For a fixed $L$, the error correction capability of the ENCRUST and the L-ENCRUST is given as,
\begin{equation}
\rho_0=\frac{L-M}{\alpha_2}. \label{ecc}
\end{equation}
We perform $l_1$ minimization for error recovery, which requires $\alpha_2$ to be in the range of four and six.{The range of four to six are empirical values widely used in compressive sensing practice, which guarantees the signal recovery with high probability \cite{ec2}.} Theoretically, when there is no quantization error, the reconstructed signal quality of the ENCRUST and the L-ENCRUST get better as increase in $M$ as shown in Fig. \ref{quant} (the red curve). However, in practice there are quantization errors. Though the reconstructed signal quality first gets better with increase in $M$ due to less lossy compression,  the signal quality stops increasing but starts to fall when $M$ exceeds for example 126 for L-ENCRUST. This is because the error correction capability is decreasing with $M$ increasing for a fixed $L$, according to Eq. \ref{ecc}.  The effect of the quantization for the ENCRUST and L-ENCRUST is shown in Fig. \ref{quant}.  
It can be observed that in general the reconstructed signal quality  of the L-ENCRUST is better as compared to that of the ENCRUST. This is because the error recovery matrix, $\mathbf{ A}$, for ENCRUST is non-sparse which results in more quantization errors while for L-ENCRUST the matrix $\mathbf{ A}$ is sparse. 
\begin{figure}[htbp]
	\centering
	\includegraphics [width=.9\linewidth]{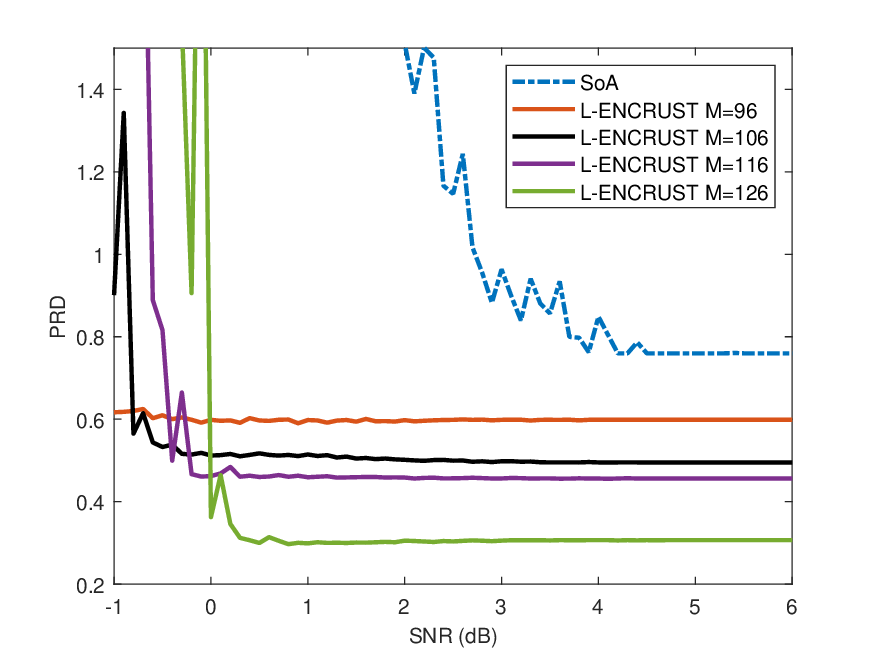}
	\caption{PRD performance of an ECG signal for $N=256$ and $L=168$ at various
		values of SNR and $M$ for IEEE 802.15.4.}
	\label{ErrorPer}
\end{figure} 

\begin{figure}[htbp]
	\centering
	\includegraphics [width=.9\linewidth]{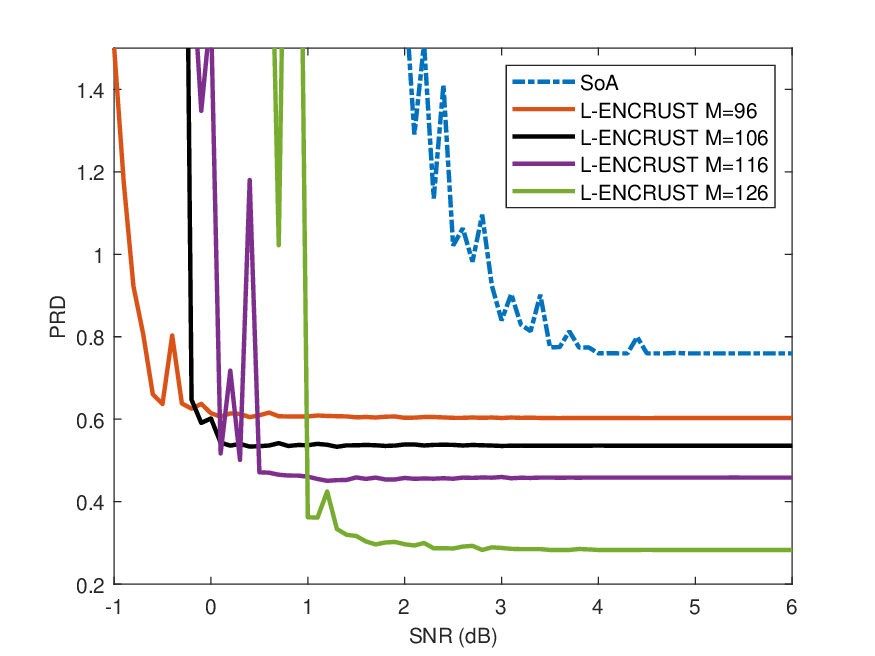}
	\caption{PRD performance of an ECG signal for $N=256$ and $L=150$ at various
		values of SNR and $M$ for IEEE 802.15.4.}
	\label{ErrorPer_LBB}
\end{figure} 

We simulate ECG signal transmission using the SoA solution, ENCRUST and L-ENCRUST under IEEE 802.15.4 settings. For the SoA solution with signal length $N=256$, the number of bits after applying HAAR on ECG signal are 1536, i.e., with compression ratio of 1536/(256*11) and  after applying Hamming(7,4) code on the HAAR compressed signal the total number of bits is 2688. Note that AES encryption is performed in the counter mode on the encoded signal. For fair comparison we take the same bit budget as of the SoA solution for L-ENCRUST  i.e., $M=96$ and $L=168$, each sample is in 16 bits. Note that since in the L-ENCRUST the projection matrix construction using the matrix $\mathbf{ A}$ is used for error recovery, the matrix $\mathbf{ A}$ should be full rank. We observe that for $\text{iv}=0$x$ffff$, $N=256$, $L=168$ and $d=15$, the matrices constructed using Algorithm \ref{algoSm} are full rank for the simulated values of $M$. 

After applying the L-ENCRUST on a block of ECG signals the following steps are taken. First, IEEE 802.15.4 frames are generated with payload of 102 Bytes. After that the bits of each frames are spreaded to chips with two samples per chip, i.e., with spreading factor 2. Finally, chips are transmitted using offset quadrature phase shift keying modulation. To have a particular signal to noise ratio (SNR) of the received signal, the modulated signal is added with white Gaussian noise. The IEEE 802.15.4 protocol is modified in the following way. If there are errors in the received frame payload but not in header, then the retransmission request is not sent. However, if there is error in received packet header, then the retransmission request  is sent, as L-ENCRUST only provides error recovery to the payload bits and leaves the IEEE 802.15.4 frame header untouched. The reconstructed ECG signal quality with changing SNR for different values of  $M$ is shown in Fig. \ref{ErrorPer}. It can be observed that the reconstructed signal quality of the L-ENCRUST with channel error and quantization error is better than that of the SoA solution, particularly at low SNR.  We also simulate the case of the L-ENCRUST for various values of $M$ and $L=150$, which makes L-ENCRUST have $10\%$ lower bit budget compared to the SoA solution, as shown in Fig. \ref{ErrorPer_LBB}. It can be observed that the L-ENCRUST scheme still outperforms  the SoA solution even with a lower bit budget.

\begin{figure}[htbp]
	\centering
	\subfloat[\label{effi2}]{\includegraphics[width=.5\linewidth]{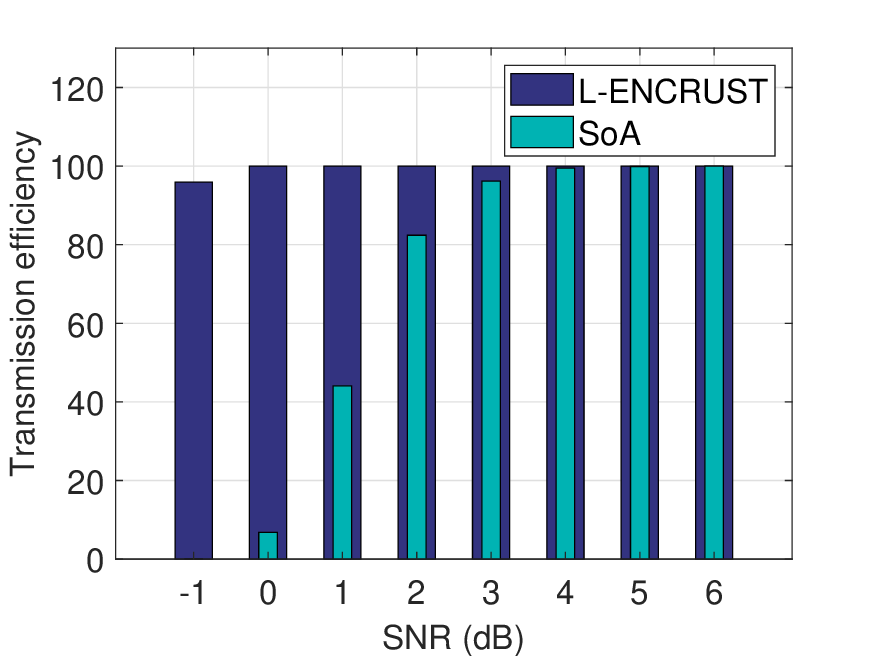}}
	\subfloat[\label{effi3}]{\includegraphics[width=.5\linewidth]{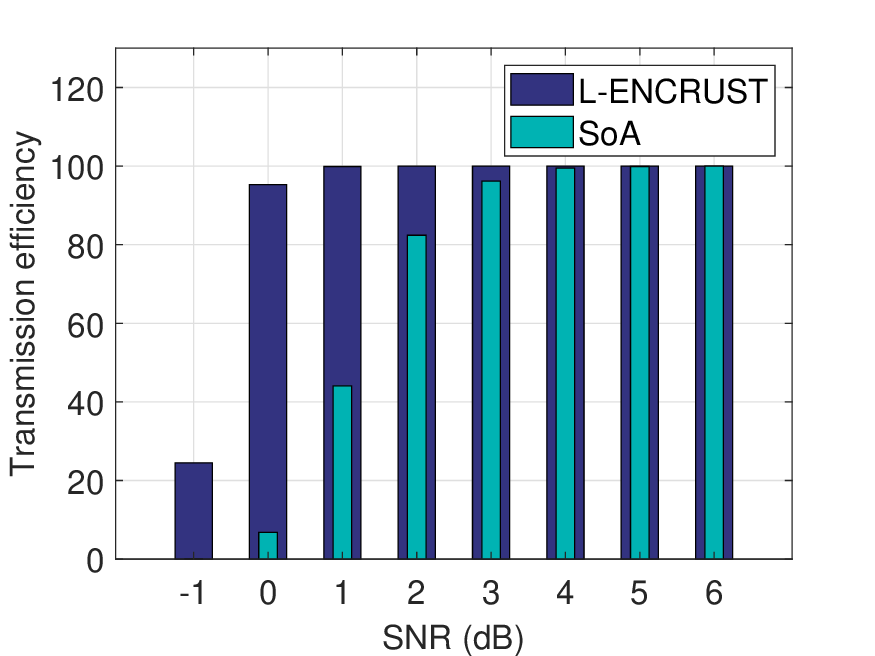}}\\
	\subfloat[\label{effi}]{\includegraphics[width=.5\linewidth]{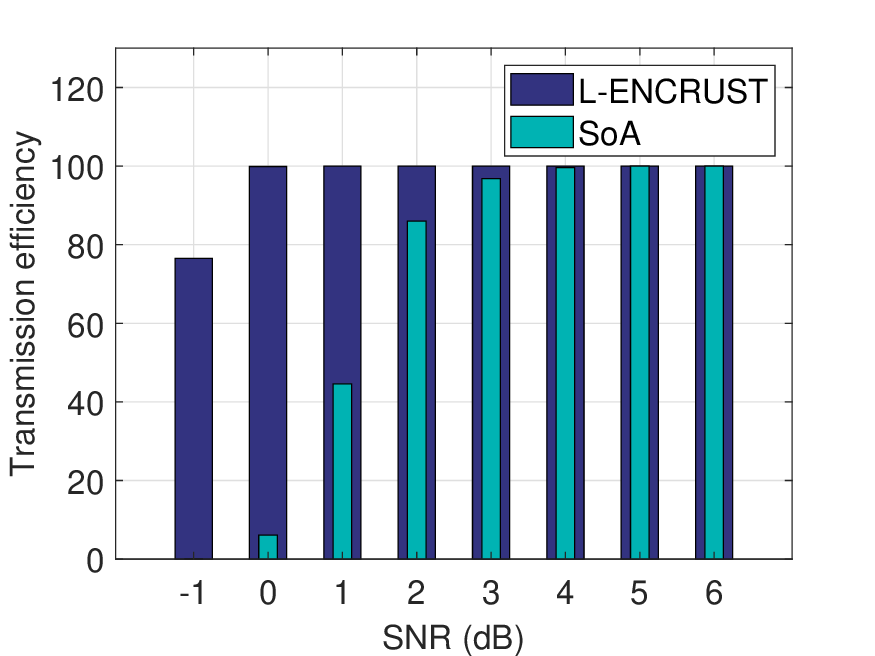}}
	\subfloat[\label{effi1}]{\includegraphics[width=.5\linewidth]{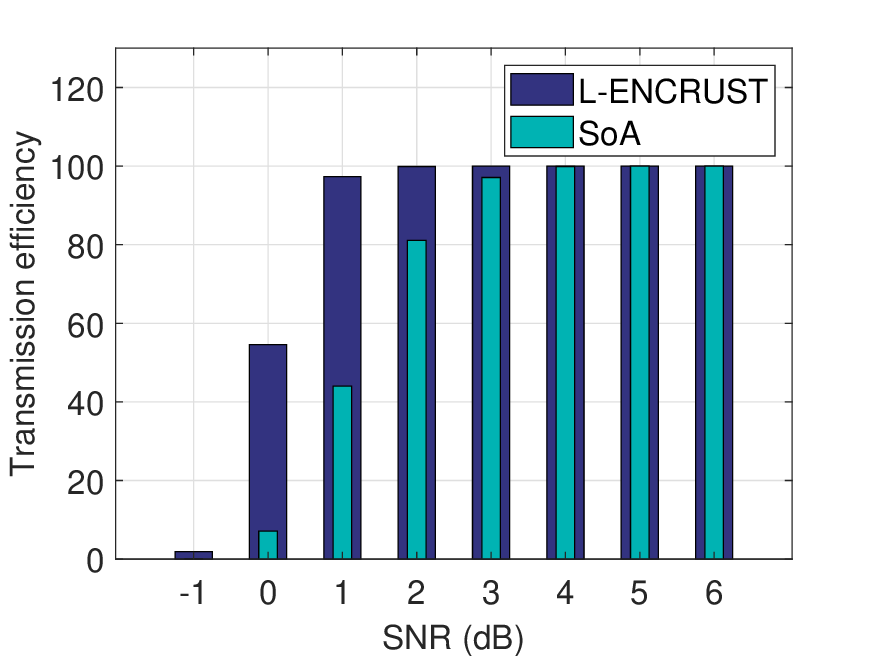}}
	\caption{ Comparison of the transmission efficiency of the SoA solution and the L-ENCRUST scheme for various values of SNR (a) $L=168$, $M=96$. (b) $L=168$, $M=126$. (c)$L=150$, $M=96$. (d) $L=150$, $M=126$. }
	\label{ovEffi}
\end{figure}

We also evaluate the transmission efficiency of the L-ENCRUST scheme and the SoA solution. The transmission is regarded as successful if the PRD of reconstructed signal is below 1, otherwise, retransmission is required. The transmission efficiency of the L-ENCRUST and the SoA solution for the transmission failure probability, $P_f$, is calculated as $(1-P_f)*100$. The simulation is run for 1000 times. Transmission efficiency of the L-ENCRUST scheme and the SoA solution is shown in Fig. \ref{ovEffi}. It can be observed that at SNR equal to $-1$ dB, it is not possible  to complete data transmission due to too high retransmission failure probability leading to extremely low transmission efficiency.  Whereas, L-ENCRUST can achieve transmission efficiency of $95\%$  and $76\%$ for the case of  $M=96$ and $L=168$, and $M=96$ and $L=150$, respectively.  Transmission efficiency of the L-ENCRUST reaches $100\%$ at  SNR of $0$ dB for $L=168$ and $1$ dB SNR for $L=150$, whereas for the SoA solution the similar performance is achieved at  SNR of $5$ dB. Note that Fig. \ref{effi2} and \ref{effi3} show the simulation results under the condition that L-ENCRUST and the SoA solution have the same bit budget, while
Fig. \ref{effi} and \ref{effi1}  show the simulation results under the condition that L-ENCRUST uses $10\%$ fewer bits as compared to the SoA solution. Comparing Fig. \ref{effi2} and \ref{effi3} or similarly comparing Fig. \ref{effi} and \ref{effi1}, it shows higher value $M$ leads to lower transmission efficiency. As we see from Fig. \ref{ErrorPer} and \ref{ErrorPer_LBB}, in general high $M$ can improve signal reconstruction quality. Therefore, there is a clearly trade-off between reconstructed signal quality and transmission efficiency. From the application's point of view, as long as the reconstructed signal can meet the required signal quality, for example, in this case PRD below 1, it is not necessary to continue increasing $M$ as it does not benefit from the overall system efficiency.  Nevertheless, the L-ENCRUST scheme outperforms  the SoA solution even for higher values of $M$.

\subsection{Prototyping and Energy Consumption Measurement} 
\begin{figure}[hthp]
	\centering
	\subfloat[]{\includegraphics[width=.9\linewidth]{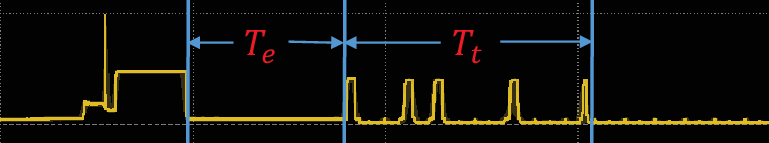}}\\
	\subfloat[]{\includegraphics[width=.9\linewidth]{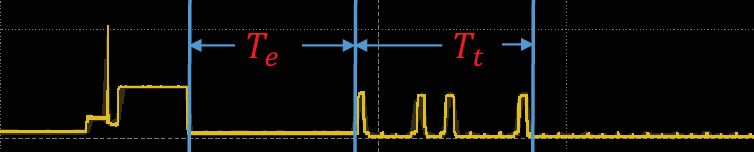}}\\
	\subfloat[]{\includegraphics[width=.9\linewidth]{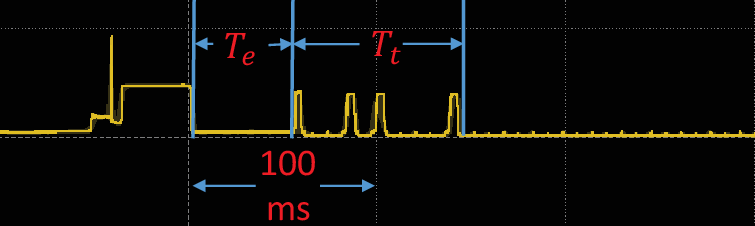}}
	\caption{Time marking for energy measurement. (a) SoA. (b) ENCRUST. (c) L-ENCRUST}
	\label{TimeMark}
	
\end{figure}

In this subsection, memory footprint, execution time, and energy consumption are measured for the SoA solution, ENCRUST, and L-ENCRUST. These schemes are implemented  using a resource-constrained hardware, TelosB mote, and cross-platform operating system for IoT devices, Contiki-NG \cite{NG}.  TelosB mote uses microcontroller TI MSP430F1611,  with 10 KB RAM 10 KB and 48 KB flash memory.  Since the SoA solution uses AES for information secrecy, the in-build software implementation of AES is used. Contiki-NG has {an} option for AES, which can be enabled by setting a flag. For data transmission, we use Contiki-NG NullNet as it is sufficient for our experiments and can keep the protocol stack simple. Frames are transmitted in a sequential manner in broadcast mode without waiting for acknowledgment.  To measure energy consumption we use $10$-$\Omega$ resistor in  series to the mote and voltage across the resister is measured using Analog Discovery2 oscilloscope at sampling rate of 1 MHz.  The transceiver CC2420 is kept at the turnoff state while executing the codes for encoding process and it is turned on just before the transmission starts. To measure energy consumption, the voltage across the $10$-$\Omega$ resister is captured using Analog discovery oscilloscope. The captured voltage waveforms for the SoA solution, ENCRUST, and L-ENCRUST are shown in Fig. \ref{TimeMark}. We use the notation, $T_e$,  for the execution time and $T_t$, for transmission time.

\begin{figure}[htbp]
	\centering
	\includegraphics [width=.7\linewidth]{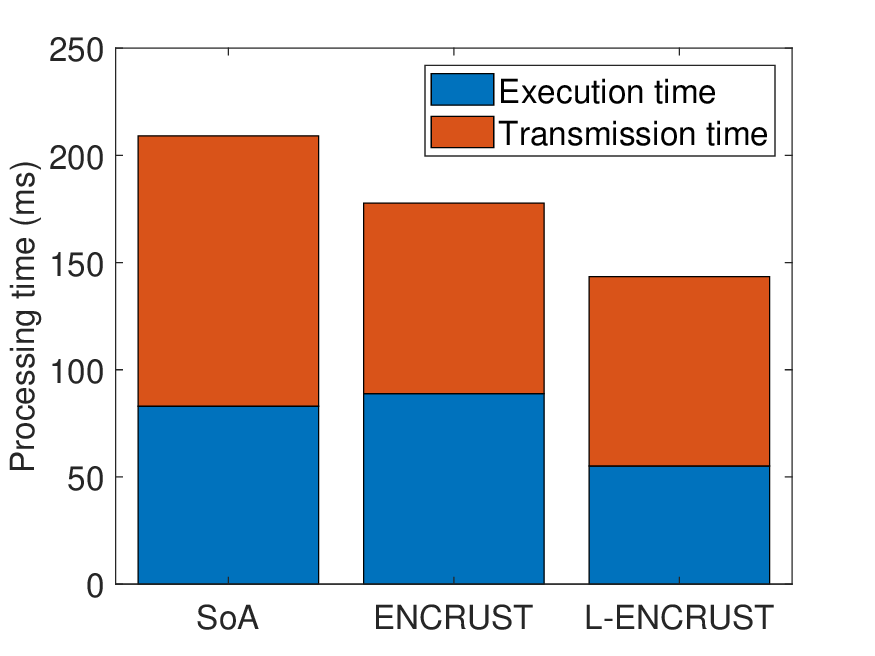}
	\caption{Processing time breakdown of the  SoA solution, ENCRUST, and L-ENCRUST schemes. }
	\label{ProTime}
\end{figure} 
\begin{figure}[htbp]
	\centering
	\includegraphics [width=.7\linewidth]{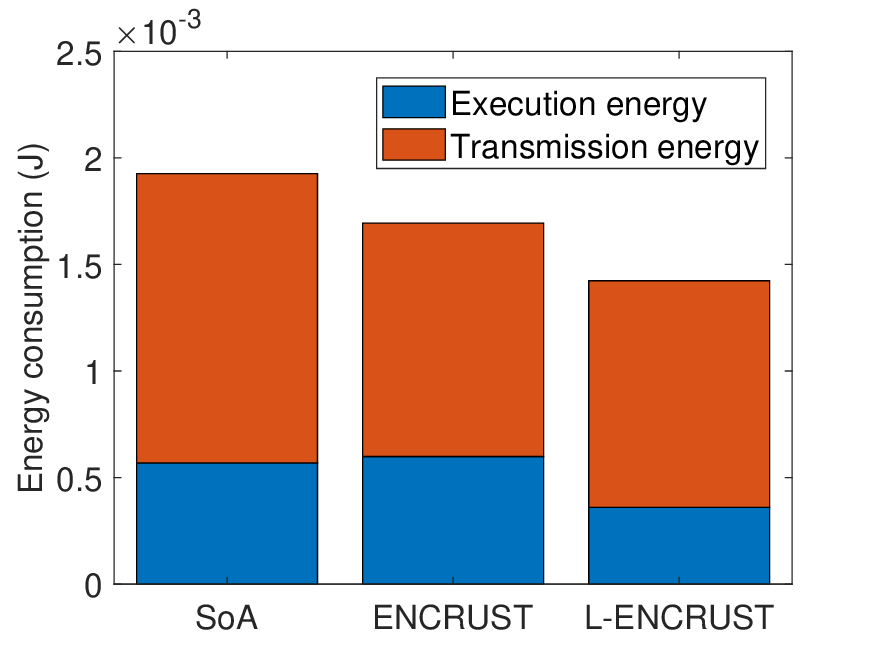}
	\caption{Energy breakdown of the  SoA solution, ENCRUST, and L-ENCRUST schemes. }
	\label{ProEnergy}
\end{figure} 
\begin{table}[]
	\caption{Comparison of the memory overhead in bytes for the  SoA solution, ENCRUST, and L-ENCRUST schemes. }
	\label{memFoot}
	\centering
	\begin{tabular}{l c c c }
		
		& \textbf{SoA}	 & \textbf{ENCRUST}  & \textbf{L-ENCRUST} \\ \toprule[.5mm]
		\multicolumn{1}{l}{\begin{tabular}[l]{@{} l@{}}\textbf{Memory}\\ \textbf{overhead} (Bytes)  \end{tabular}}& 1342	& 860 &    728 \\ \bottomrule[.1mm]
	\end{tabular}
\end{table}


\subsubsection{Memory Footprint}
The memory footprint is measured to evaluate the code storage size of different solutions.  Memory overhead for the SoA solution, ENCRUST, and L-ENCRUST are shown in Table \ref{memFoot}. The memory overhead of a particular scheme is calculated by  subtracting the basic memory footprint of Contiki-NG operating system from the total memory footprint for that scheme. From Table \ref{memFoot}, it can be observed that the memory overheads for the L-ENCRUST and ENCRUST scheme  are reduced by $45\%$ and $35\%$,  as compared to that of the SoA solution. Due to the lower memory footprints, ENCRUST and L-ENCRUST are more suitable to  resource-constrained IoT devices.


\begin{table}[]
	\caption{Performance measurement of the L-ENCRUST scheme for various values of $M$ and $L=150$.}
	\label{variSS}
	\centering
	\begin{tabular}{l c c c  c}
		
		$M$	& $96$ & $106$ & $116$ & $126$ \\ \toprule[.5mm]
		\multicolumn{1}{l}{\begin{tabular}[l]{@{}l@{}}\textbf{Execution} \rule{0pt}{2.6ex}\\ \textbf{time}, $T_e$ (ms)\end{tabular}} & 53    & 58     & 61     & 64     \\ 
		\multicolumn{1}{l}{\begin{tabular}[l]{@{}l@{}}\textbf{Execution energy} \rule{0pt}{2.6ex}\\ \textbf{consumption} (mJ) \end{tabular}}   & 0.38  & 0.40   & 0.43   & 0.50   \\ \bottomrule[.1mm]
	\end{tabular}
\end{table}

\begin{table*}[]
	
	\caption{Performance measurement of the various possible scenarios of the SoA solution and the L-ENCRUST.}
	\label{variS}
	\centering
	\begin{tabular}{ccc|cc|cc}
		
		& \multicolumn{2}{c}{\begin{tabular}[c]{@{}c@{}}\textbf{Compression \&} \\ \textbf{information secrecy}\end{tabular}} & \multicolumn{2}{c}{\begin{tabular}[c]{@{}c@{}}\textbf{Error recovery \&} \\ \textbf{information secrecy}\end{tabular}} & \multicolumn{2}{c}{\begin{tabular}[c]{@{}c@{}}\textbf{Compression, error recovery} \\ \textbf{ \& information secrecy}\end{tabular}} \\ 
		\multicolumn{1}{c}{}                                                                   & SoA \rule{0pt}{3.6ex}                                          &  \multicolumn{1}{c}{\begin{tabular}[c]{@{}c@{}}{L-ENCRUST }\\ $M=96$   \end{tabular}}                                      & SoA                                            & \multicolumn{1}{c}{\begin{tabular}[c]{@{}c@{}}{L-ENCRUST }\\ $L=308$   \end{tabular}}                                   & SoA                                                   & \multicolumn{1}{c}{\begin{tabular}[c]{@{}c@{}}{L-ENCRUST }\\ $M=96$ and $L=150$   \end{tabular}}                                                 \\ \toprule[.5mm]
		\multicolumn{1}{l}{\begin{tabular}[l]{@{}l@{}}\textbf{Memory}\\  \textbf{overhead} (Bytes)\end{tabular}} & 1116                                          & 340                                               & 1338                                           & 330                                                 & 1342                                                  & 728                                                        \\ \hline
		\multicolumn{1}{l}{\begin{tabular}[l]{@{}l@{}}\textbf{Execution}\rule{0pt}{2.6ex}\\ \textbf{time}, $T_e$ (ms)\end{tabular}}      & 43                                            & 24                                                & 165                                            & 79                                                  & 83                                                    & 53                                                         \\ \hline
		\multicolumn{1}{l}{\begin{tabular}[l]{@{}l@{}}\textbf{Transmission} \rule{0pt}{2.6ex}\\ \textbf{time}, $T_t$ (ms)\end{tabular}}  & 50                                            & 53                                                & 225                                            & 224                                                 & 126                                                   & 90                                                         \\ \hline
		\multicolumn{1}{l}{\begin{tabular}[l]{@{}l@{}}\textbf{Total time}\rule{0pt}{2.6ex}\\ $T_e+T_t$ (ms)\end{tabular}}                                                    & 93                                            & 77                                                & 390                                            & 303                                                 & 209                                                   & 143                                                        \\ \hline
		\multicolumn{1}{l}{\begin{tabular}[l]{@{}l@{}} \textbf{Energy} \rule{0pt}{2.6ex}\\ \textbf{consumption} (mJ)\end{tabular}} & 1                                             & 0.8                                               & 3.8                                            & 2.9                                                 & 1.9                                                   & 1.4                                                        \\ \bottomrule[.1mm]
	\end{tabular}
\end{table*}

\subsubsection{Energy Consumption and Processing Time} ENCRUST and L-ENCRUST are implemented for $N=256$, $M=96$, and $L=150$, and the SoA solution is implemented for signal length $N=256$, as described in Subsection \ref{ComMe}.  The processing time for a scheme is divided into execution time and transmission time. The total processing time and its breakdown for the SoA solution, ENCRUST and L-ENCRUST are shown in Fig. \ref{ProTime}.  It can be observed that the total processing time of the ENCRUST and L-ENCRUST is smaller than that of the  SoA solution. ENCRUST and L-ENCRUST reduce the processing time by $14\%$ and $31\%$ as compared to the SoA solution, respectively.

The energy consumptions for the SoA solution, ENCRUST, and L-ENCRUST  are measured with the same parameters as described above. The total energy consumption is the sum of energy consumed for execution of a scheme and energy consumed for data transmission. In Fig. \ref{ProEnergy}, it shows that ENCRUST and L-ENCRUST achieve $12\%$ and $26\%$ reduction in the total energy consumption as compared to the SoA solution, respectively. We also study the increased cost on execution energy and execution time by increasing the value of $M$. The measurement results of the L-ENCRUST scheme is shown in Table \ref{variSS}.  The increase in the execution time and energy is not linear, which is likely due to the
optimization performed by Contiki compiler. Note that the transmission time and transmission energy remain the same, since $L$ is fixed to $150$.

\begin{figure}[]
	\centering
	\includegraphics [width=.8\linewidth]{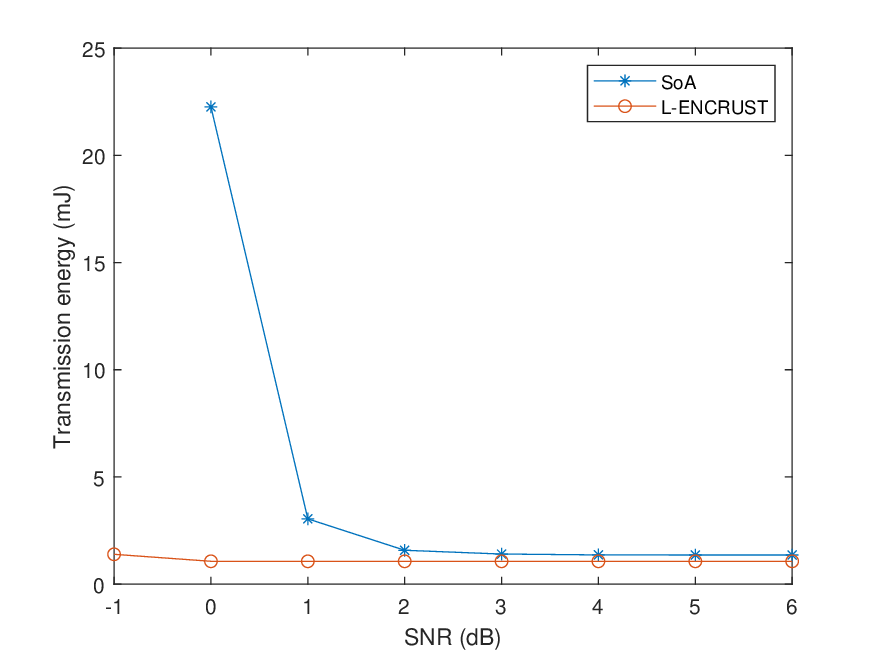}
	\caption{Total transmission energy consumption of the SoA solution and L-ENCRUST at $M=96$, $L=150$ and $N=256$ at various SNR values.  }
	\label{ProEnergy1}
\end{figure}

Till now the performance measurements are performed without considering the effect of the channel errors. In Fig. \ref{ProEnergy1}, we show the effect of the channel errors on the transmission energy of a block of ECG signal with length $256$.

Transmission efficiency for $L=150$ and $M=96$ are shown in Fig. \ref{ovEffi}. We demonstrate the effect of the channel noise on the transmission energy in Fig. \ref{ProEnergy1}. It can be observed that at SNR of $-1$ dB data communication is not possible using the SoA solution and at SNR of  $0$ dB the SoA solution consumes 20x energy compared to the L-ENCRUST
scheme. It can also be observed that the L-ENCRUST has overall lower transmission energy consumption  than the SoA solution for low SNR range.

Besides L-ENCRUST achieves less processing time, better energy efficiency, higher transmission efficiency and smaller memory footprint, it provides a great agility in offering data compression, error recovery and information secrecy depending on the need of applications and channel conditions. For example, when Eq. \ref{n3} of L-ENCRUST encoding is written as $\mathbf{y}_i=\mathbf{Bx}_i+\mathbf{r}_i$, it provides only compression and information secrecy. Similarly, when Eq. \ref{n3} of L-ENCRUST encoding is written as $\mathbf{y}_i=\mathbf{Ax}_i+\mathbf{r}_i$, it provides only error recovery and information secrecy. In Table \ref{variS}, we show the experimental results of the possible configuration scenarios of L-ENCRUST in comparison with the SoA solution. Note that the dimensions of $\mathbf{ A}$ and $\mathbf{ B}$ are determined based on the equivalent values used in the SoA solutions. It can be
observed that L-ENCRUST scheme outperforms the SoA solutions for all possible scenarios in terms of memory overhead, total processing time, and energy consumption.

\section{Conclusion}
In this paper, we design and implement a working prototype of the  theoretical  ENCRUST scheme, which can simultaneously perform compression, error recovery, and information secrecy.  We present  construction algorithms for compression matrix and error recovery matrix to achieve  energy-efficient operations in resource-constrained devices. We also purpose a new lightweight variant of the ENCRUST, named as L-ENCRUST. Security analysis is performed for both ENCRUST and L-ENCRUST, which are shown to be secure against ciphertext-only attack, known-plaintext attack, and chosen-plaintext attack.  Prototypes of the state-of-the-art solution, ENCRUST, and L-ENCRUST are realized in a real resource-constrained IoT device. {The prototype of ENCRUST and L-ENCRUST are tested on the TelosB mote, showing that they can reduce energy consumption and memory overhead by 12\% and 26\%, and 35\% and 45\%, respectively, compared to the state-of-the-art solution.
	We list the possible future directions for the ENCRUST and L-ENCRUST:
	\begin{itemize}
		\item Further optimization of the ENCRUST and L-ENCRUST schemes to improve their energy efficiency and memory overhead.
		\item Integration of the ENCRUST and L-ENCRUST schemes into real-world applications to test their practicality and usefulness.
		\item Comparison of the ENCRUST and L-ENCRUST schemes with other state-of-the-art solutions for compression, error recovery, and information secrecy.
	\end{itemize}
}

\end{document}